\documentclass{article}
\usepackage{float}
\usepackage[utf8]{inputenc}
\usepackage[caption = false]{subfig}
\usepackage{spconf,amsmath,graphicx}
\usepackage{xcolor}
\usepackage{hyperref}
\usepackage{lipsum,multicol}

\usepackage{amsfonts}
\usepackage{ulem}
\usepackage{tikz}
\usepackage{stmaryrd}
\usepackage{amsthm}
\usepackage{enumitem}
\usepackage{multirow}
\usetikzlibrary{automata, arrows.meta, positioning}

\usepackage{enumitem}
\let\OLDthebibliography\thebibliography
\renewcommand\thebibliography[1]{
  \OLDthebibliography{#1}
  \setlength{\parskip}{0pt}
  \setlength{\itemsep}{0pt plus 0.2ex}
}

\graphicspath{ {img/} }
\title{\vspace*{-\baselineskip}Adaptive Sampling for Storage of Progressive Images on DNA\vspace*{-0.5\baselineskip}\thanks{\tiny This work was funded by the European Union's Horizon research and innovation programme projects Glaciation (Grant No. 101070141), and ANR PEPR program MoleculArxiv}}
\name{Xavier Pic$^{\star}$, Nimesh Pinnamaneni$^{\star\star}$, Raja Appuswamy$^{\star}$}
\address{$^{\star}$ EURECOM, Data Science Department, Sophia Antipolis, France\\$^{\star\star}$ Helixworks Technologies, Ltd., MTU Campus, Bishopstown, Cork, Ireland\vspace*{-2.5\baselineskip}}

\begin{document}

\maketitle
% \vspace*{-\baselineskip}
\begin{abstract}
% \vspace*{-0.5\baselineskip}
The short lifespan of traditional data storage media, coupled with an exponential increase in storage demand, has made long-term archival a fundamental problem in the data storage industry and beyond. Consequently, researchers are looking for innovative media solutions that can store data over long time periods at a very low cost. DNA molecules, with their high density, long lifespan, and low energy needs, have emerged as a viable alternative to digital data archival. However, current DNA data storage technologies are facing challenges with respect to cost and reliability. Thus, coding rate and error robustness are critical to scale DNA storage and make it technologically and economically achievable. Moreover, the molecules of DNA that encode different files are often located in the same oligo pool. Without random access solutions at the oligo level, it is very impractical to decode a specific file from these mixed pools, as all oligos  need to first be sequenced and decoded before a target file can be retrieved, which greatly deteriorates the read cost.

This paper introduces a solution to efficiently encode and store images into DNA molecules, that aims at reducing the read cost necessary to retrieve a resolution-reduced version of an image. This image storage system is based on the Progressive Decoding Functionality of the JPEG2000 codec but can be adapted to any conventional progressive codec. Each resolution layer is encoded into a set of oligos using the JPEG DNA VM codec, a DNA-based coder that aims at retrieving a file with a high reliability. Depending on the desired resolution to be read, the set of oligos as well as the portion of the oligos to be sequenced and decoded are adjusted accordingly. These oligos will be selected at sequencing time, with the help of the adaptive sampling method provided by the Nanopore sequencers, making it a PCR-free random access solution. 
\vspace*{-0.5\baselineskip}
\end{abstract}
\begin{keywords}
% \vspace*{-0.5\baselineskip}
DNA data storage, JPEG 2000, progressive, random access, JPEG DNA VM
\vspace*{-0.5\baselineskip}
\end{keywords}
% \vspace{-1\baselineskip}
\vspace*{-0.5\baselineskip}
\section{Introduction}
\vspace*{-0.5\baselineskip}

The data storage industry is facing unprecedented challenges as the demand for storage continues to increase at an exponential rate. Conventional storage devices, have fundamental durability and density limitations that make long-term data storage infeasible. 
% As a result, s
Synthetic DNA molecules, has emerged as an alternative for overcoming these challenges \cite{Church}, mainly due to their long lifespan, high density and low energy needs. 

A classic solution for storing data onto DNA molecules is composed of both biochemical and computational processes. The two main biochemical processes are synthesis, to create DNA molecules with the desired sequences of nucleotides, and sequencing, to retrieve the sequence of nucleotides that it contains. The main computational process is the DNA-adapted codec, that transforms data into quaternary sequences for synthesis, and back during sequencing. The codec also compresses the data to reduce synthesis and sequencing cost. Prior work on DNA storage has focused on optimizing both biochemical and computational processes with the goal of transforming DNA into a reliable digital storage medium despite errors that arise at various points in the reading and writing pipeline. However, current DNA storage solutions are not designed to support an important type of access mode that is triggered when images are stored in DNA--the adaptive selection of the decoded image resolution.

Adaptive resolution selection is used in several scenarios. For instance, depending on the client used to visualize the image, a social media application might choose to display a low resolution versions of images on resource constrained mobile devices, versus high resolution versions on desktops. Supporting adaptive resolution selection effectively requires the use of both efficient image coding techniques that can compress images to enable progressive decoding, and random access support from underlying storage media to be able to retrieve selective portions of an image depending on resolution requirements. 
%Historically, large image collections have been stored on conventional magnetic data storage media, like hard disk drives or tapes, that have supported random access. 
%Current data storage technologies almost always support random access, that enables the rapid access of specific data. When transmitting multimedia, this technology allows for the tansmitted data to be adapted to the client (dektop or mobile). 
Prior work on DNA data storage has demonstrated the use of Polymerase Chain Reaction (PCR) for enabling such random access. However, another alternative, based on the Read Until function of the Nanopore sequencers, can help retrieve oligos with specific reference sequences, without doing any primer-specific PCR augmentation. For now, such random access has only been applied to arbitrary binary files and neither have been used in the context of image storage for features like adaptive selection of decoded resolution.

In this paper, we bridge this gap by presenting a novel approach that brings together progressive image coding with Read Until functionality of Nanopore sequencers to support the adaptive selection of resolution for large image collections stored in DNA, at a very low read cost compared to existing solutions. 
\vspace*{-1\baselineskip}
\section{Context}
\vspace*{-0.5\baselineskip}
\subsection{DNA-adapted coding}
\vspace*{-0.5\baselineskip}
The research activities in the field of DNA data storage have been rapidly developing over the past few years. In this section, we provide an overview of a few pioneering approaches, both in the general case for DNA storage, and more specifically, for the case of image storage in DNA.
The recent survey~\cite{dna-storage-survey} makes detailed comparisons of various approaches that focus on storing generic, binary data using DNA. In 2012, Church et al. \cite{Church}, introduced an approach to enable large-scale encoding and decoding of data into synthetic DNA molecules. This paper not only gave a methodology for encoding and decoding data into DNA molecules, but also identified the main coding constraints that have to be respected when encoding data into DNA molecules. In 2013, Goldman et al.\cite{Goldman} provided one of the first encoders capable of respecting some of these constraints. The provided entropy coder allowed for the encoding of any file into DNA-like data. In 2015, Grass et al.\cite{Grass} introduced the first error correction codes into a DNA data storage solution. This error correction mechanism aims at robustifying the whole storage process against the errors (substitutions, insertions, deletions) occurring because of the biochemical operations. Following this, other works introduced other error correction solutions \cite{Erlich,Yazdi,Aeon,Blawat, GCNSA, Adaptive-coding, Reconstruction-DNA}, by adding redundancy to binary data to detect and correct errors. Finally, random access solutions aiming to enable selective access to binary data stored in DNA and improve read cost have also been investigated \cite{organick,CMOSS}.

Besides the solutions encoding generic binary data, DNA-adapted image codecs also emerged. One of the first solutions, Dimopoulou et al.\cite{Dimopoulou} developed a JPEG-based image coder adapted to DNA data storage. Similarly, in \cite{IMG-DNA}, the authors introduce a DNA-adapted coding scheme based on JPEG. Pic el al. used an improved DNA-adapted entropy coder\cite{SFC4} to increase the performance of this JPEG-based DNA-adapted image coder. Lazzarotto et al.\cite{EPFL-Raptor} developed the JPEG DNA VM software that encodes data with a system based on Raptor codes\cite{Raptor-codes}. It was output by the JPEG DNA ad-hoc group as a verification model for later developments. Learning-based image compression methods have also been studied~\cite{dna-qlc,eurecom}, where the authors use a variational autoencoder to compress the image into a latent space that is later encoded into DNA. In \cite{seongjun}, the authors leverage pixel domain representation to reconstruct the images with better compression performance.
\vspace*{-1\baselineskip}
\subsection{Random Access in DNA Data Storage}
%This part has to be reduced a little bit
\vspace*{-0.5\baselineskip}
% Random access in DNA storage systems refers to the selective retrieval of specific data files from a heterogeneous pool of DNA oligonucleotides without the need to sequence the entire dataset. This capability is essential for practical DNA storage implementations, as bulk sequencing becomes prohibitively expensive and time-consuming as data volumes scale to terabytes or petabytes. 

The majority of random access methods involves the incorporation of unique address sequences or physical handles that enable selective identification and isolation of target DNA strands. These addressing schemes must satisfy several critical requirements: orthogonality to prevent cross-reactivity, thermodynamic stability under operational conditions, and compatibility with downstream sequencing and decoding processes \cite{organick,(7)Tomek2019,(8)ElShaikh2022,(9)Lin2020,(10)Subramanian2018,(11)Gowri2024}.
\vspace*{-0.5\baselineskip}
\subsection{PCR-based Random Access}
\vspace*{-0.25\baselineskip}
PCR-based random access represents the first scalable approach to selective data retrieval in DNA storage systems. This method employs primer sequences appended to oligos, enabling selective amplification of target files through PCR. The University of Washington and Microsoft collaboration demonstrated the first large-scale implementation, successfully retrieving individual files from a pool containing over 13 million DNA oligos, encoding 200 megabytes of data \cite{organick}.

The system architecture requires careful primer design to ensure orthogonality and prevent non-specific amplification \cite{organick, (10)Subramanian2018,(13)Zhang2013,(58)Booeshaghi2023}. Primer libraries must satisfy stringent thermodynamic constraints, including controlled GC content (40-60\%), annealing temperatures above 58°C, and minimal formation of secondary structures \cite{(10)Subramanian2018,(13)Zhang2013}.
%Recent advances have enabled the design of validated primer sets with coding capacities exceeding 13,000 unique targets while maintaining cross-talk rates below 0.11\% \cite{(10)Subramanian2018}.

PCR-based systems demonstrate excellent scalability, with theoretical capacity for $8.98\times 10^{21}$ addressable targets and a system capacity of 65.8 ZB \cite{(7)Tomek2019}. However, several limitations constrain their practical implementation. The amplification process is inherently destructive, requiring complete consumption of the original DNA pool for each access operation \cite{organick}. Access times range from several hours due to thermal cycling requirements, and elaborate workflows such as emulsion PCR \cite{organick,(7)Tomek2019} and micro-encapsulation techniques employed to overcome amplification bias that can skew representation of different sequences \cite{(15)Bogels2023}. Additionally, using primer sequences, particularly nested or hierarchal primers \cite{organick, (7)Tomek2019} within an oligo requires 20-60nt overhead, taking away from nucleotide allocation to a data payload in a 200nt oligo, which increases the cost of synthesis.
\vspace*{-0.5\baselineskip}
\subsection{Silica Encapsulation with Surface Labelling}
\vspace*{-0.25\baselineskip}
The silica encapsulation approach addresses several limitations of PCR-based systems by physically protecting DNA within impervious silica capsules which are surface-labeled with single-stranded DNA barcodes corresponding to addresses or file metadata \cite{(16)Banal2021}. Unlike PCR retrieval, FACS-based sorting of silica-encapsulated DNA allows for the non-destructive selection of target files without amplification, avoids bias, and decouples address labels from payload strands, improving net data density.
%This method also enables Boolean logic queries, such as "president AND 18th century," to retrieve specific files based on multiple criteria.

The encapsulation process utilises sol-gel chemistry to create stable silica matrices around DNA molecules \cite{(16)Banal2021}. Experimental research demonstrated that DNA encapsulated in silica maintains structural integrity and remains sequenceable after accelerated aging equivalent to 2000 years in central Europe \cite{Grass}. The silica matrix protects against degradation from UV radiation, oxidation, and hydrolysis \cite{(19)Kapusuz2017}.

Fluorescence-activated cell sorting (FACS) enables precise selection of target capsules from complex mixtures \cite{(16)Banal2021}. The system achieves selection sensitivity of one in $10^{6}$ files per optical channel, with the capability to scale to $10^{6}N$ files using common commercial FAS systems which offer up to $N = 17$ optical channels \cite{(16)Banal2021}. FACS operates by detecting oligonucleotide probes with fluorescent labels hybridised to barcode sequences on the surface of a silica capsule, allowing rapid identification and physical separation of target capsules while leaving the rest of the pool intact for future access \cite{(24)BioRadFACS2025}.

Encapsulation preserves the integrity of the pool for repeated queries, and the silica matrix confers long-term environmental protection against hydrolysis, oxidation, and UV damage \cite{(16)Banal2021,Grass}. However, this approach involves elaborate workflows for sol-gel encapsulation and surface labeling, introduces latency due to chemical deprotection steps required to release DNA from capsules before sequencing, and relies on sophisticated and costly instrumentation such as FACS. Furthermore, throughput can be limited by the hybridization kinetics of fluorescent probes and the sorting speed, and designing probes that avoid cross reactivity in large libraries adds further complexity and expense \cite{(16)Banal2021,(19)Kapusuz2017,(24)BioRadFACS2025}.
\vspace*{-0.5\baselineskip}
\subsection{Adaptive Sampling with Nanopore}
\vspace*{-0.25\baselineskip}
Adaptive Sampling exploits the real-time control capabilities of nanopore sequencing to implement PCR-free random access. As each DNA strand begins translocation, buffered current signals are basecalled and aligned against user-provided address or metadata sequences; if a match is detected, sequencing proceeds, otherwise voltage reversal ejects the strand back into the pool. Roman Sokolovskii et al. demonstrated that this approach can selectively enrich target subsets from a heterogeneous DNA pool, dynamically switch targets mid-run, and employ library recovery protocols so that non-target strands remain available for future retrievals \cite{(59)Sokolovskii2024}. The core mechanism requires buffering and processing on the order of $ \sim 400\,\mathrm{nt}$ of signal plus basecalling and alignment latency (800nt total) before a keep-or-eject decision can be made \cite{(60)NanoporeAdaptiveSampling}, meaning that sufficiently long DNA constructs ($\ge 1000$--$1500\,\mathrm{nt}$) are needed for reliable early alignment against short address motifs.

Compared to PCR-based random access, adaptive sampling inherently avoids amplification bias and pool depletion, whereas adaptive sampling ejects non-matching strands intact, preserving library diversity and enabling reuse. Dynamic target switching during a run allows multiple files to be retrieved from a single aliquot without additional preparation steps \cite{(59)Sokolovskii2024}. Relative to silica encapsulation with FACS-based sorting, adaptive sampling removes the need for sol-gel encapsulation workflows, surface labelling, chemical deprotection, and specialized sorting instrumentation; selection occurs inline during sequencing via in-read alignment, simplifying operations and reducing preparative overhead. Although adaptive sampling's efficiency diminishes for very rare targets in large pools without prior enrichment, and its need for long constructs requires assembly or similar workarounds, its non-destructive, dynamic, and amplification-free characteristics make it a compelling alternative or complement to PCR- and silica-based random access methods.

% \vspace{-1\baselineskip}
\vspace*{-0.5\baselineskip}
% {\color{teal}
\section{Proposed method}
\vspace*{-0.75\baselineskip}
\subsection{Oligo Structure and Payload}
\begin{figure}
    \centering
    \includegraphics[width=0.85\linewidth]{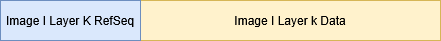}
\vspace*{-0.75\baselineskip}
    \caption{\small Oligo structure for the Adaptive Sampling sequencing mode}
\vspace*{-1\baselineskip}
    \label{fig:oligo-format}
\end{figure}

\vspace*{-0.5\baselineskip}
The adaptive sampling method of Random Access relies on the presence of reference sequences, located at the beginning of oligos. These sequences are used as indexes for the following long payload, appended to the end of the reference (Fig. \ref{fig:oligo-format}). This long payload is built by synthesizing and ligating shorter oligos together, using adapters.
%\subsections{Payload design}

Our image coding solution is based on the JPEG2000 codec in its progressive coding mode. The codec outputs a bitstream organized in $N_{levels}$ resolution layers of increasing sizes. This bitstream is sliced into a set of $N_{levels}$ binary files, corresponding to the different layers. Each resolution binary file is encoded into a pool of DNA-adapted short payload oligos using the JPEG DNA VM codec. Each pool will then be separately synthesized and prepared for adaptive sampling. In this design, the reference sequence previously mentioned indexes a specific image resolution layer.
\vspace*{-0.75\baselineskip}
\subsection{Reference sequences and dataset preparation}
\vspace*{-0.5\baselineskip}
The reference sequences used for each resolution are designed during the preparation of the oligos. A reference sequence dictionary is built to associate any resolution level of an image with the reference sequence that was used for the preparation of its pool of oligos. Once all the resolution levels of all the images in the dataset have been prepared, they can be merged into a single general pool of oligos, the dataset. For each resolution layer of each image, a reference sequence will be stored in the dictionary, and provided back, should the image be decoded. 
% {\color{purple}
It is important to note here that in the case where a thumbnail-based access system as the one presented in \cite{DSP25} is adapted for adaptive sampling, a short DNA sequence could be appended to the end of each thumbnail oligo, and used as a seed for the design of the reference sequences, removing the need to store them externally.
% }
\vspace*{-0.75\baselineskip}
\subsection{PCR-free Random Access and image retrieval}
\begin{figure*}
    \centering
% \vspace*{-0.75\baselineskip}
    \includegraphics[width=0.9\linewidth]{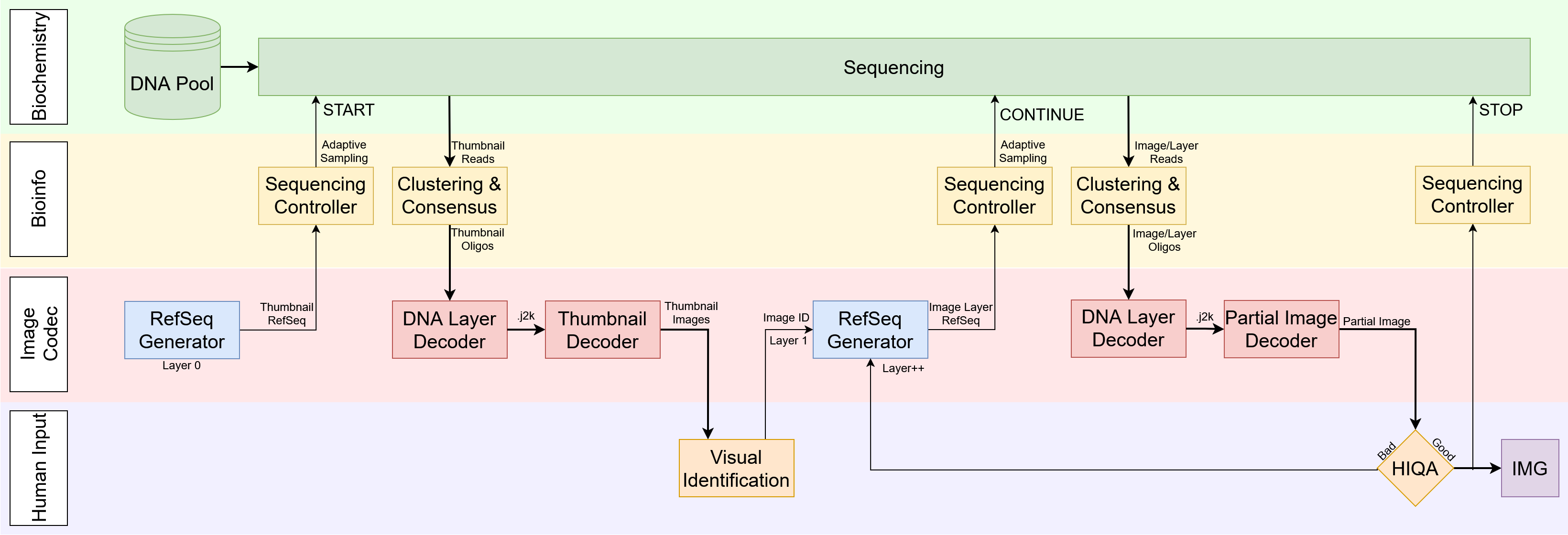}
\vspace*{-0.5\baselineskip}
    \caption{\small Thumbnail-based decoding workflow of the Adaptive Sampling based DNA image codec}
\vspace*{-1\baselineskip}
    \label{fig:decoding-wrkflw}
\end{figure*}

\vspace*{-0.5\baselineskip}
Progressive decoding was used in the past to mitigate the constraints imposed by bad internet connections when exchanging images: while the data was being transmitted, the first layers could progressively be decoded, to obtain a degraded preview of the desired image. In the same manner, in DNA data storage, improvements on the sequencing cost could be leveraged from the use of such a progressive decoding technique.
An image can be progressively retrieved from the general oligo pool, by looking, in the dictionary of reference sequences, for the sequences of the different resolution layers of that specific image. The different layers of the image are retrieved starting with the lower resolutions. 
% {\color{purple}
During the sequencing process, the image codec will successively input different reference sequences to the Nanopore sequencer, that will update the parameter of the running Read Until process. The current reference sequence that is input to the Read Until process will allow the sequencer to only sequence the oligos starting with this input, and reject back into the general oligo pool the other ones.
% }
The general image codec (Fig.\ref{fig:decoding-wrkflw}) is used to progressively select, sequence and decode these different encoded layers, to little by little refine the final desired image. When a satisfactory resolution quality is output by the codec, an early stopping instruction can be sent to the sequencer, to avoid sequencing unnecessary data. 
% }

% {\color{red} 
% \section{Proposed Method}
% \vspace*{-\baselineskip}
% % \vspace*{-1.25\baselineskip}
% \subsection{Description of image coding solution the solution}
% \vspace*{-0.6\baselineskip}
% \vspace{-1\baselineskip}
\vspace*{-0.75\baselineskip}
\section{Experimental Results}
\vspace*{-0.5\baselineskip}
The proposed codec is primarily designed to decrease the cost necessary to retrieve an image from a pool of oligos. 
The first step to measure this cost is to identify the real use case for image retrieval. In DNA coding systems that do not implement any random access at the oligo level, the cost of reading an image is equal to the number of oligo reads necessary to read the whole dataset. For this type of coder, Equation (\ref{eq:def1}) can represent the read-cost necessary for the retrieval of an image. On the other hand, an image retrieval process that integrates random access or progressive decoding in its core coding system can improve on the read cost in Equation (\ref{eq:def1}). Firstly, if progressive decoding is enabled, the read cost can be limited to Equation (\ref{eq:def2}), since only the first layers need to be decoded. 
% {\color{purple}Secondly, if a thumbnail-based access designed as in \cite{DSP25} is also enabled, read-cost can be represented as Equation (\ref{eq:def3}). In this last case, the double sum is replaced by two sums, the first part represents the cost to decode the thumbnails, while the second denotes the cost to read the rest of the desired image.}
\vspace*{-0.25\baselineskip}
\begin{equation}
    R_c(I, K) = \frac{\sum_{i=0}^{N_{images}}\sum_{k=0}^{N_{levels}}nucs(i, k)}{input\_image\_pixels}
\label{eq:def1}
\end{equation}
\vspace*{-0.25\baselineskip}
\begin{equation}
    R_{c\_pd}(I, K) = \frac{\sum_{i=0}^{N_{images}}\sum_{k=0}^{K}nucs(i, k)}{input\_image\_pixels}
\label{eq:def2}
\end{equation}
% \vspace*{-0.75\baselineskip}
% \begin{equation}
%     R_{c\_ra}(I, K) = \frac{\sum_{i=0}^{N_{images}}nucs(i, 0) + \sum_{k=1}^{K}nucs(I,k)}{input\_image\_pixels}
% \label{eq:def3}
% \end{equation}
% \vspace*{-0.25\baselineskip}
    The $nucs(i,k)$ value represents the number of nucleotides to sequence to be able to decode a layer:
% \vspace*{-0.75\baselineskip}
\begin{equation}
\vspace*{-0.25\baselineskip}    
    nucs(i,k) = coverage(i,k) \times number\_oligos(i,k)
\vspace*{-0.25\baselineskip}    
\end{equation}
With this, we can define a read-cost gain that measures the improvements provided by our progressive decoding approach:
% With this, we can define a read-cost gain as in two different ways, depending on whether or not Random Access is enabled (in both cases, Progressive decoding is used) as:
% \vspace*{-0.5\baselineskip}
\begin{equation}
    G_{pd}(I, K) = \frac{R_c(I, K)}{R_{c\_pd}(I, K)}\\
\vspace*{-0.5\baselineskip}
\end{equation}
% \begin{equation}
%     G_{ra}(I, K) = \frac{R_c(I, K)}{R_{c\_ra}(I, K)}\\
% \vspace*{-0.5\baselineskip}
% \end{equation}
\vspace*{-0.75\baselineskip}
\subsection{Performance evaluation}
\vspace*{-0.25\baselineskip}
The performance of any DNA encoding method can be evaluated with respect to a series of metrics such as RD-curves, reading cost and writing cost. As our work primarily focuses on progressive decoding (PD) 
% {\color{purple}and thumbnail-based image random access (RA)}
, the main metric we focus on in our study is the reading cost. We especially study the evolution of the reading cost for a given encoded image, as we read through each resolution layer. The image is encoded into a series of resolution layers
% {,\color{purple}the smallest one being the thumbnail}
. 
The resolution layers each divide the size of the image by a factor of 2 in each dimension (4 in total).
% Hence the estimated gain on the reading cost should approach the estimated values in Table \ref{tab:read_costs}. Finally the RD curve in Figure \ref{} shows the evolution of the reconstruction quality of the image while decoding the different layers. 
The results presented here were obtained by encoding 5 images of the kodak\footnote{\url{https://r0k.us/graphics/kodak/}} dataset, each with 3 resolution levels. 
The oligos had data blocks of length 148. During synthesis, these data blocks will be ligated together and reference sequences specific to the image and resolution level will be added to each ligated molecule, for Oligo Random Access with adaptive sampling. 
A theoretical read-cost gain was observed (Table \ref{tab:read_costs}) that depicts the improvements provided by progressive decoding against a similar coding method that does not provide any adaptive read-cost reduction such as the selection of resolution. 
% {\color{purple}lines 3 and 5 of }

% The top part of Table \ref{tab:read_costs} represents the gains obtained when only progressive decoding is enabled. 
In the results shown in Table \ref{tab:read_costs}, we measured the number of oligos necessary to decode each image until a certain resolution level, and averaged the results over all the images. In these conditions, the progressive decoder provides gains of up to 7.5$\times$, when only the initial layer is targeted. This gain quickly decreases if more layers are targeted, and if all the layers are targeted, no gain can be leveraged from PD, because all oligos need to be sequenced. These gains, though, heavily depend on the dimensions of the chosen resolution layers: smaller layers will leverage better gains, at the cost of more distorted images.
% {\color{purple}The bottom part of Table \ref{tab:read_costs} represents the gains obtained when both PD and RA are enabled. In these gains, the cost necessary to retrieve the thumbnail is included, as described in the second member of the sum in Equation (\ref{eq:def3}). The random access process further improves the read cost, especially in layers with better resolution where the gain is multiple times larger than the one measured without random access. This gain highly depends on the number of images encoded in the pool of DNA molecules, and on the size of the different resolution layers.}
% Additionally, Figures \ref{fig:sub_a} and \ref{fig:sub_b} depict the evolution of the reconstruction quality (PSNR here) of the image, as a  function of the read cost (which depends on the resolution layers that are selected).
% With the reduction factors that we used in the progressive coding parameters, it is possible to decode a degraded version of the image for a fraction of the read-cost necessary to decode the whole picture. For instance, the thumbnail shown in Figure \ref{fig:sub_c} has a PSNR of 18.3dB\footnote{ Additional data can be found here: \\{\url{https://gitlab.eurecom.fr/pic/jp2dnaprogressiveres}}}(to compute the PSNR, the reduced image was resized to the original image's dimensions with a bi-cubic interpolation). The last resolution level (Fig. \ref{fig:sub_d}), in contrast, shows a PSNR of 52dB. Further, the thumbnail is of good enough quality to be used as a visual reference in the random access process (Fig. \ref{fig:secondary_primer}). 

The gains obtained are orthogonal to improvements in coding performance and read cost that can be achieved by adjusting the encoding options of both the JPEG DNA VM and JPEG2000. 
The JPEG DNA VM software parameters can be adjusted differently for each resolution layer, especially the redundancy, so that lower resolution layers are better protected against errors.

% \begin{table}[]
%     \centering
%     \resizebox{6cm}{!}{
%     \begin{tabular}{cc||c|c|c|}
%         \cline{2-5}
%         \multicolumn{1}{c|}{}&Layer & $L_0$ & $L_1$ & $L_2$ \\
%         \cline{2-5}
%         \noalign{\vskip-1\tabcolsep \vskip\arrayrulewidth \vskip-2.2\doublerulesep}
% \\ \hline

%         \multicolumn{1}{|c||}{\multirow{3}{*}{PD}}&\# Oligos & 802 & 2229 & 6208 \\
%         \cline{2-5}
%         \multicolumn{1}{|c||}{}&Theoretical $G_{pd}$& \multirow{2}{*}{7.74} & \multirow{2}{*}{2.78} & \multirow{2}{*}{1}  \\
%         \multicolumn{1}{|c||}{}&Read-cost gain &&&\\
%         \hline
%         \hline
%         % Layer &$L_0$& $L_1$ & $L_2$ & $L_3$ & $L_4$ \\
%         % \hline
%         \multicolumn{1}{|c||}{\multirow{3}{*}{RA \& PD}}&\# Oligos & 802 & 1087 & 1883 \\
%         \cline{2-5}
%         \multicolumn{1}{|c||}{}&Theoretical $G_{ra}$& \multirow{2}{*}{7.74} & \multirow{2}{*}{5.71} & \multirow{2}{*}{3.30} \\      
%         \multicolumn{1}{|c||}{}&Read-cost gain &&&\\
%         \hline
%     \end{tabular}}
% \vspace*{-0.25\baselineskip} 
%     \caption{Average read-cost gains $G_{pd}$ and $G_{ra}$ for each target resolution level $L_{k}$, averaged over all the selected images.}
% \vspace*{-0.5\baselineskip}
%     \label{tab:read_costs}
% \end{table}

\begin{table}[]
    \centering
    \resizebox{5cm}{!}{
    \begin{tabular}{|c|c|c|c|}\hline
        Layer & $L_0$ & $L_1$ & $L_2$\\\hline
        \# Oligos & 802 & 2229 & 6208\\\hline
        \multicolumn{1}{|c|}{Theoretical $G_{pd}$}& \multirow{2}{*}{7.74} & \multirow{2}{*}{5.71} & \multirow{2}{*}{3.30}  \\
        \multicolumn{1}{|c|}{Read-cost gain} &&&\\\hline
    \end{tabular}}
    \vspace*{-0.75\baselineskip}
    \caption{\small Average read-cost gains $G_{pd}$ for each target resolution level $L_{k}$, averaged over all the selected images.}
    \vspace*{-1\baselineskip}
    \label{tab:read_costs}
\end{table}
\vspace{-1\baselineskip}
\vspace*{-0.4\baselineskip}
\section{Conclusion}
\vspace*{-0.75\baselineskip}
In this paper, we introduced a progressive image codec, adapted to DNA data storage, that improves the read cost necessary to retrieve an image, through progressive decoding and thumbnail-based random access. 
This novel codec allowed us to leverage read cost that are multiple times lower than the ones necessary to decode the data without progressive decoding or thumbnail-based access.
These theoretical results motivated us to place an order for the synthesis of those oligos, and conduct real wet-lab experiments in the future months.
\vspace*{-1\baselineskip}
\small
\bibliographystyle{IEEEbib}
\bibliography{biblio}

\end{document}